\documentclass{acm_proc_article-sp}
\usepackage[english]{babel}
\usepackage{graphicx}
\usepackage{fancyhdr}
\begin{document}

\conferenceinfo{}{Bloomberg Data for Good Exchange 2016, NY, USA}

%\title{Tourism and Events Analytics with Mobile Phone Records: a Case Study in Andorra}

\title{Analysis of Tourism Dynamics and Special Events through Mobile Phone Metadata }

\numberofauthors{6}
\author{
	\alignauthor
	Yan Leng\\
	\affaddr{MIT Media Lab}\\
	\affaddr{Cambridge, MA}\\
	\email{yleng@mit.edu}
	\and
	\alignauthor
	Alejandro Noriega\\
	\affaddr{MIT Media Lab}\\
	\affaddr{Cambridge, MA}\\
	\email{noriega@mit.edu}
	\and
	\alignauthor Alex 'Sandy' Pentland\\
	\affaddr{MIT Media Lab}\\
	\affaddr{Cambridge, MA}\\
	\email{pentland@mit.edu}
	\and
	\alignauthor Ira Winder\\
	\affaddr{MIT Media Lab}\\
	\affaddr{Cambridge, MA}\\
	\email{jiw@mit.edu}
	\and
	\alignauthor Nina Lutz\\
	\affaddr{MIT Media Lab}\\
	\affaddr{Cambridge, MA}\\
	\email{nlutz@mit.edu}
	\and
	\alignauthor Luis Alonso\\
	\affaddr{MIT Media Lab}\\
	\affaddr{Cambridge, MA}\\
	\email{alonsolp@mit.edu}	
}\maketitle

\begin{abstract}
Tourism has been an increasingly important factor in global economy, society and environment, accounting for a significant share of GDP and labor force. Policy and research on tourism traditionally rely on surveys and economic datasets, which are based on small samples and depict tourism dynamics at low spatial and temporal granularity. Anonymous call detail records (CDRs) are a novel source of data, showing enormous potential in areas of high societal value: such as epidemics, poverty, and urban development. This study demonstrates the added value of using CDRs for the formulation, analysis and evaluation of tourism strategies, at the national and local levels. In the context of the European country of Andorra, we use CDRs to evaluate marketing strategies in tourism, understand tourists' experiences, and evaluate revenues and externalities generated by touristic events. We do this by extracting novel indicators in high spatial and temporal resolutions, such as tourist flows per country of origin, flows of new tourists, tourist revisits, tourist externalities on transportation congestion, spatial distribution, economic impact, and profiling of tourist interests. We exemplify the use of these indicators for the planning and evaluation of high impact touristic events, such as cultural festivals and sports competitions. \footnote{This paper is a draft for submission for Data for Good Exchange (2016). The improvements in synthesis, analysis and writing are yet to be implemented throughout.}
\end{abstract} 

% A category with the (mi on Systems Applications}{Miscellaneous}
%A category including the fourth, optional field follows...
%\category{D.2.8}{Software Engineering}{Metrics}[complexity measures, performance measures]

%\category{H.4}{Information Systems Applications}{Miscellaneous TODO}

%\terms{TODO}


\section{Introduction}

Tourism has been an increasingly important factor in global economy, society and environment, accounting for a significant share of GDP and labor force \cite{proencca2008tourism, unwto}. Economists and governments have been attempting to understand the contribution and impact of tourism to economy for years \cite{balaguer2002tourism, song2012tourism}. The growing importance of event tourism, an applied field devoted to understanding and improving tourism through events, have resulted in substantial global competition and generated values for both public and private sectors \cite{getz2016event}. Research on events and tourism planning, operation and marketing have grown exponentially since 2008, which involves accommodation, attraction, transport and ancillary services\cite{Getz2016593}. The success of tourism events expand the potential of tourism and the capacity of the tourism area to step beyond a narrow leisure-based tourism \cite{Getz2016593}. 

% The complexity of the research lies in the spatial and temporal heterogeneity inherent in the supply and demand \cite{getz2016event}.

Most tourism and events analytics literatures are based on surveys, interviews, observations or focus group studies \cite{Getz2016593, crowther2015review, balaguer2000tourism, dupeyras2013indicators}. Pettersson (2009) \cite{pettersson2009event} studied the spatial and temporal nature of event experiences through interviews, participant observation and photography at a major sporting event in Sweden, contributing to a better understanding of how visitors interact with the event setting and with each other, and help build theory on experiences.  Li (2008) \cite{li2008systematic} conducted two-phase online surveys with six-month interval to analyze first- time and repeat visitors via demographic characteristics, travel planning behavior, pre and post-trip congruency in travel activity preferences, and post-trip evaluation on first-time and repeated visitors. However, these studies are small in size, unrepresentative, low in spatial-temporal resolutions, and unable to measure travel experiences.

In recent years, big data, mobile sensors and social media, generated many incredible opportunities for its supposed capacity to provide answers for questions related to travelers' behaviors and experiences \cite{baggiobig}. Applying big data in tourism has the following advantages: larger reliability than self-reported data and intentions; easier to cross-reference with other data based on the geo-spatial information; more knowledge about the industry's target market producted by the customers themselves. Birenboim (2016) \cite{birenboim2016new} tracked 25 students with an mobile application, which triggers up to three micro-surveys an hour for subjective evaluation of momentary sense of crowdedness and sense of security. Wood (2013) \cite{wood2013using} combined field investigation and Flickr data, showing that Flickr data can be used to understand what elements of nature attract tourists and how this would alter visiting behaviors. Bassolas (2016) \cite{bassolas2016touristic} analyzed the attractiveness of tourists sites using geopositioned twitter. 

Large-scale geo-located Information and Communication Technology, such as Call Detail Records collected from mobile phones, created new ways to understand the interconnectedness among travelers, social events, urban infrastructures \cite{bassolas2016touristic}. It has been the most salient source of information, eliciting large-scale patterns of human mobility \cite{wheatmand_noriega}. CDRs have been applied various fields, such as mobility modeling \cite{alhasoun2014city}, epidemics \cite{friassimulation}, disaster response \cite{blondel2015survey}. However, this large-scale and longitudinal data source has seldom been used in tourism analytics. In this study, we demonstrate how to utilize mobile phone data to evaluate tourism marketing strategies, understand tourists' experiences, evaluate revenues and externalities generated by tourism. 

This paper contributes and adds value to tourism planning and event analytics by novel indicators extracted from mobile phone records.  With higher spatial-temporal resolutions comparing with self-reported surveys, CDRs reveals more insights and improved knowledge into various aspects of tourism industry. We extract various indicators, including tourists flows per country of origin, dynamics of new tourists, flows of revisits, tourists externalities due to congestion, spatial distribution, economic impacts and profiling of tourists interests. These indicators help other industries to enhance services they offer and the management. The study can be replicated in other regions and sectors aiding better decision-makings. In the context of Andorra, we use CDRs to evaluate marketing strategies in tourism, understand tourists' experiences, and evaluate revenues and externalities generated by touristic events. We exemplify the use of these indicators for the planning and evaluation of high impact touristic events in 2015, such as cultural festivals and sports competitions. 

The remaining of the paper is structured as follows: Section \ref{data} introduces the context of Andorra and the data used in this study. Section \ref{method} describes the novel indicators extracted mobile phone records. Section \ref{res} summaries the performances of events. Finally, section \ref{future} concludes this study and discusses future works. 
%\newpage

\section{Andorra and Data}
\label{data}
In this section, we briefly introduce the context of Andorra and the data used to conduct the study.

\subsection{Andorra}
\begin{figure}
	\includegraphics[width=1\linewidth,natwidth=2226,natheight=1962]{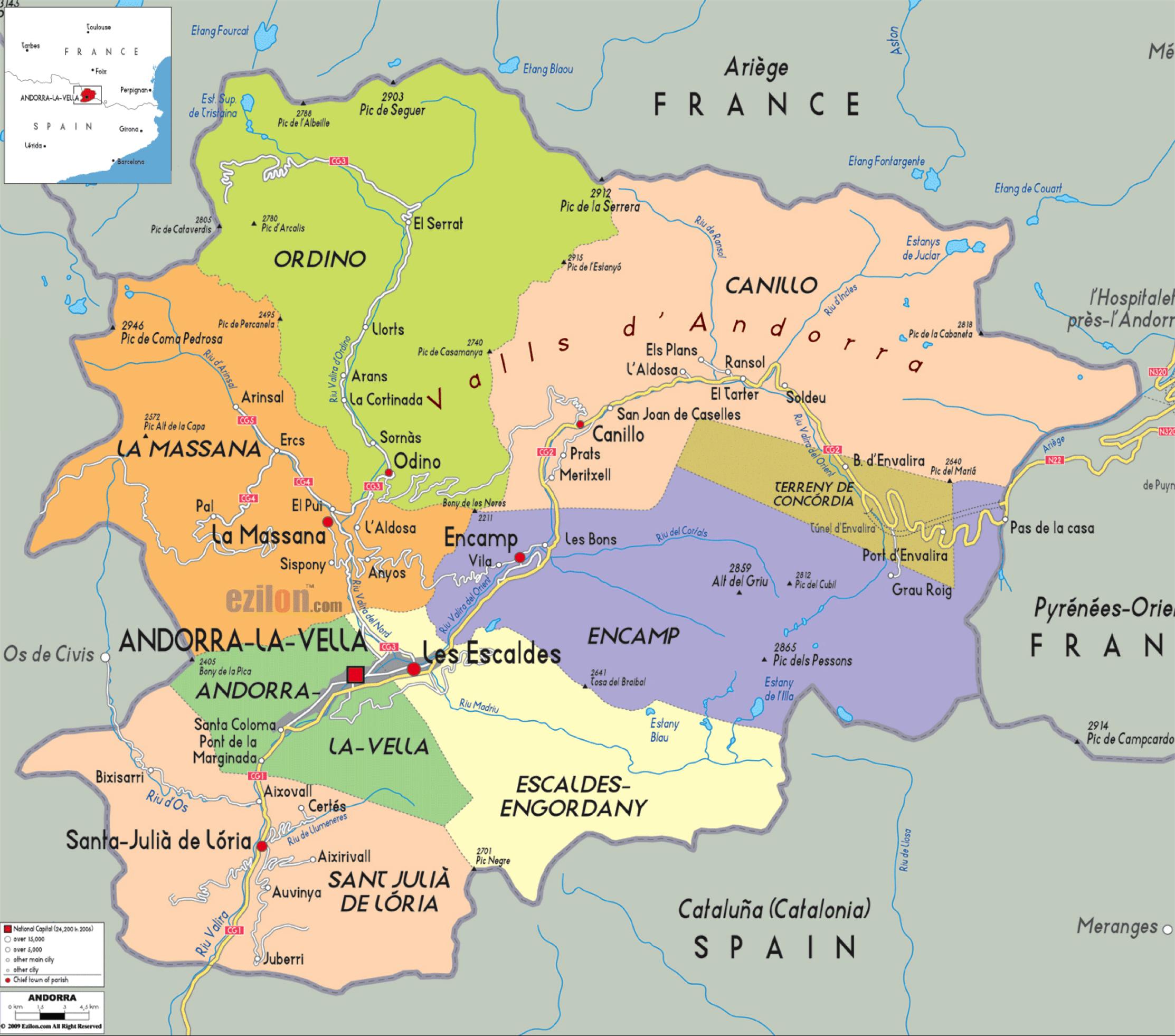}
	\caption{Map of Andorra}
	\label{andorra}
\end{figure}

Andorra is a small country in Europe covering an area of 468 $km^2$, situated near France and Spain, as shown in Figure \ref{andorra}. The population of Andorra is 85,000, with an annual of 10.2 million international visitors \cite{andorra}. Andorra is a country heavily relying on tourism and famous for skiing and duty-free shopping, with 80\% of GDP comes from tourism \cite{gdp}.

%
%\subsection{Tourism events}
%A country heavily dependent on tourism, Andorra government and tourism department hold various events to attract visitors and investments, boost the tourism sector, which turn tourism into a major contribution to improve the Andorra's economy and competitiveness \cite{getz2008event}. As a case study, we focus on several summer and winter events in Andorra.
%\begin{itemize} 
%	\item Summer events: 
%	\begin{itemize}
%		\item Storia by Cirque du Soleil: July 15 - July 20, 2015 (one week of the one-month event)
%		\item Volta als Ports d'Andorra: July 12, 2015
%		\item MTB Masters World Championships: August 23 - 27, 2015
%		\item Tour of Spain: August 31 - September 3, 2015
%		\item UCI Trial Masters World Championships: September 1 - 6, 2015
%	\end{itemize}
%	\item Winter events: 
%	\begin{itemize}
%		\item Free-ride Junior World Championship: February 7 - 8 , 2015
%		\item Speed Skiing World Championship: February 28 - March 3, 2015
%		\item Alpine Skiing European Cup Finals: March 18 - March 22, 2015
%		\item Total Fight Master of Freestyle: March 26 - April 4, 2015 
%	\end{itemize}
%\end{itemize}

\subsection{Data}

\begin{figure}
	\centering
	\includegraphics[width=1\linewidth,natwidth=831,natheight=78]{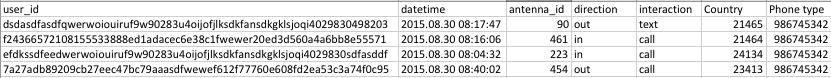}
	\caption{A snapshot of Call Detail Records}
	\label{cdr}
\end{figure}

\begin{figure}
	\includegraphics[width=1\linewidth,natwidth=671,natheight=484]{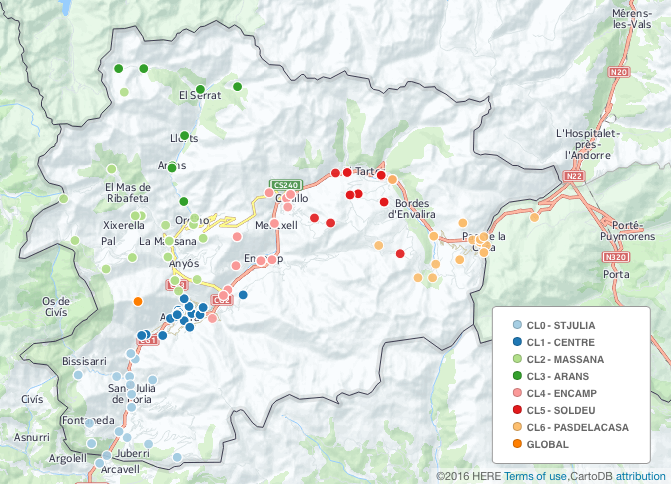}
	\caption{Cell tower distribution in Andorra}
	\label{tower}
\end{figure}

The main data source we use, as stated in section \ref{intro}, is Call Detail Records (CDRs). It is collected by Andorra telecom, originally for billing purposes. It contains metadata about call, Short Message Service (SMS) and data communications, with information on longitude, latitude, timestamp of the transaction, registry country, phone type, i.e., as shown in Figure \ref{cdr}. The CDR data used in this study was collected over a period of two years from 2014 to 2016. The distribution of cell towers are shown in Figure \ref{tower}.

Another dataset coupled with CDR is the road network, enabling the mapping from the cell-tower-based Origin-Destination matrix observed from mobile phone records to traffic flows on road links. We acquired the GIS shapefiles of road network from the Andorra transportation department. 

%\begin{figure}[h]
%	\includegraphics[width=1\linewidth]{IMG/cdr}
%	\caption{Call Detail Record}
%\end{figure}

%\begin{figure}
%	\includegraphics[width=1\linewidth]{IMG/nation.png}
%	\caption{Call Detail Record}
%\end{figure}

\section{Visualization of tourism}

\begin{figure}
	\includegraphics[width=1\linewidth,natwidth=2786,natheight=1634]{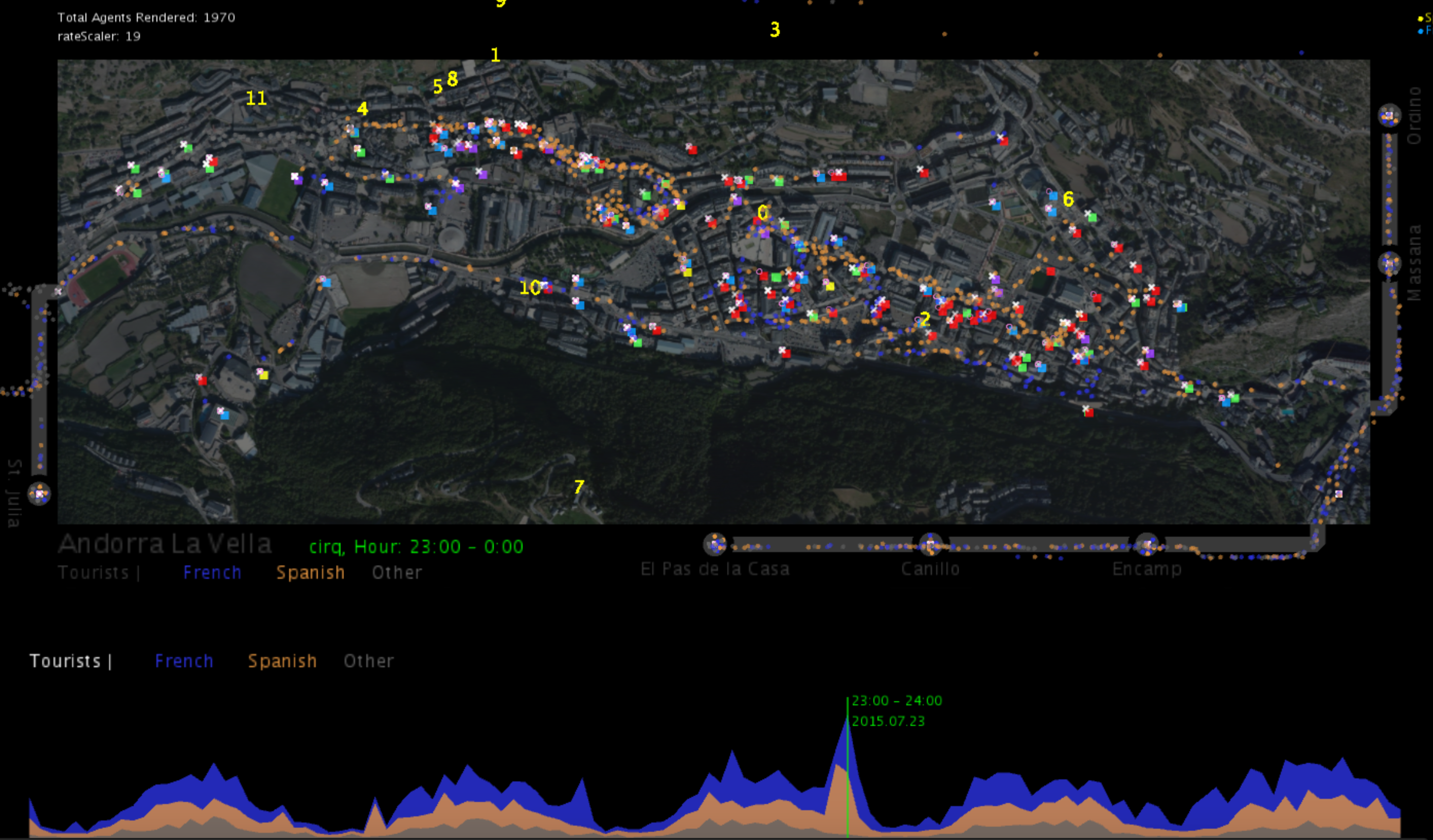}
	 Each purple and orange circle represents a French and a Spanish traveler respectively. The colored squares stand for different Points of Interests (POIs). 
	\caption{Visualization of tourists dynamics}
	\label{viz}
\end{figure}

To understand how tourists and locals interact with urban environment - transportation and commercial system - we develop a visualization platform to leverage and cross-reference different geo-spatial data. This provides an integral tool for policy makers to observe visual analytics, uncover hidden patterns and develop actionable insights. Furthermore, it make big data accessible and digestiable to narrate, engage and educate a larger audience. Figure \ref{viz} is a visualization of an event in 2015, Cirque De Soleil.

\section{Indicators mined from CDRs}
\label{method}

This section aims to describe indicators mined from CDRs, which enable decision-makers and planners to gain deep and multi-dimensional understanding in tourism performance. These indicators cover various aspects, including marketing, travel experiences, economic impacts and externalities.

% \footnote{The demo of the Spanish and the French during Cirque De Soleil can be seen at:  }

\subsection{Segmented tourists flows}
%Number of tourists and tourist days are two ways to measure tourists flows. The first one counts the total number of tourists . 
Tourist flow is a traditional indicator in tourism industry, traditionally estimated from survey. It can be measured by tourist volumes or tourist-days, which can be segmented based on country of origins directly observed from Call Detail Records. Visitors from different countries reveal different visitatoin patterns, such as the gathering regions, length of stays, interested events. Understanding the patterns across different market segmentations enables businesses to adapt offers to increase the attraction to tourists. In the context of Andorra, the main countries include Spain, France, Russia, Belgium, Portugal, and Netherlands. Spanish and French take up 90\% of the total number of tourists.

\subsection{New tourists and repeated tourists}

Marketing strategies and the performance events can be measured by new tourists and repeated tourists. Specifically, new tourist volumes reflect the marketing strategy prior to the events. Repeated tourist volumes reclect travel experiences during the event \cite{lee2009investigating, thiumsakfactors}. Correlating repeated visit rates with other factors, such as congestions, implies whether these factors are significant for travel experiences. 

New tourists are calculated as the tourists that do not visit prior to the event period. Repeated tourists are defined as tourists that revisit after the event.

\subsection{Spatial distribution}

Spatial distribution of tourists measure the spatial impact of the event during the event period. Success event generates wider impact, distributing revenues across the nation. It can be measured seperately across daytime and night since activities are different during these two periods. 

Specifically for hospitality, understanding the spatial distribution at night based on origin countries enables hotels managers to pinpoint countries to focus marketing.

\subsection{Congestion}
Congestion is the externality generated by tourism that adversely impact  local communitie, travelers' experiences and environmental system. Popular events, seemingly sucessful in attractiveness, are exposed to higher risk of congestion, which, on the other hand, become an disadvantage. Severe congestion during event indicates the need for improving tourist demand management or providing more services. 

Mobile phone data provides a dynamic monitoring of transportation system performance. The ability to track traffic flows at a granular spatio-temporal scale provides deeper understanding of tourists' travel patterns, enabling better planning and management. 

\begin{figure}
	\includegraphics[width=1\linewidth,natwidth=546,natheight=450]{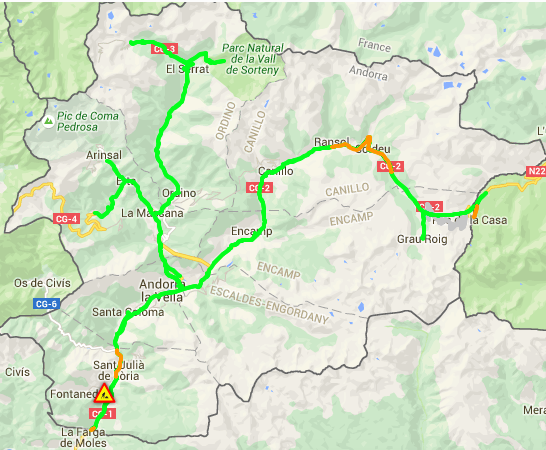}
	\caption{Andorra road network}
	\label{road}
\end{figure}

We propose a three-step method to estimate traffic flow. First, we compute the tower-to-tower Origin-Destination (O-D) matrix by aggregating the individual movements between cell towers. It is important to estimate traffic flows across temporal and spatial dimensions due to the heterogeneous variability. Second, the O-D pairs are assigned to road links, as shown in Figure \ref{road}. The last step is to scale the aggregated movements to actual vehicle trips using traffic counts as the ground truth. 
%, which is calculated in Equation \ref{flow_2}. 
%
%\begin{equation}
%\label{flow_1}
%R_i = \sum_{jk}O_jD_k
%\end{equation}
%
%\begin{equation}
%\label{flow_2}
%TC_i = R_i\times \beta_i
%\end{equation}
%
%where $i$ is the index of road link. $O_jD_k$ represents OD pair origins from $j$ and destined towards $k$. $R_i$ is the vehicle trips along road link $i$. $TC$ is the actual traffic counts. $\beta$ is the scaling factor.  
%
%In transportation, practitioners and planners model the relationships between traffic flow and travel time based on the characteristics of the road infrastructure. One of the most simplistic and widely used method is Bureau of Public Road function \cite{akcelik1991travel}, which models travel time as a function of the ratio between actual traffic volume and road maximum flow capacity, volume-over-capacity (VOC) \cite{ben1999discrete}, as shown in Equation (\ref{bpr}). 
%
%Simulated travel time as a function of traffic flow: 
%\begin{equation}
%\label{bpr}
%t_{\text{current}} = t_{\text{free-flow}} \times (1 + \alpha(V/C)^\beta)
%\end{equation}
%
%
%where $t_{\text{free-flow}}$ is the free flow travel time on the road segment. $t_{\text{current}}$ is the simulated travel time. $\Delta{t}$ is the delay caused by congestion. $\alpha$ and $\beta$ are parameters that are used to characterize the non-linear relationship between $V/C$ and $t_{current}$. According to Bureau of Pubic Roads, the default value are $\alpha=0.15$ and $\beta=4$. 

\subsection{Revenues}

Understanding the spending patterns at individual level and aggregated level enables businesses to promote offers to concentrate marketing efforts. It also helps businesses to indentify the potential markets for growth. Under the situation that no data is directly related to tourists' spendings, we come up with an approach to proxy the revenues. Revenues are positively correlated with the disposable income of tourists, which can, to some degree, be proxied by the price of the phone being used.

\begin{figure}[h]
	\includegraphics[width=\linewidth,natwidth=1350,natheight=760]{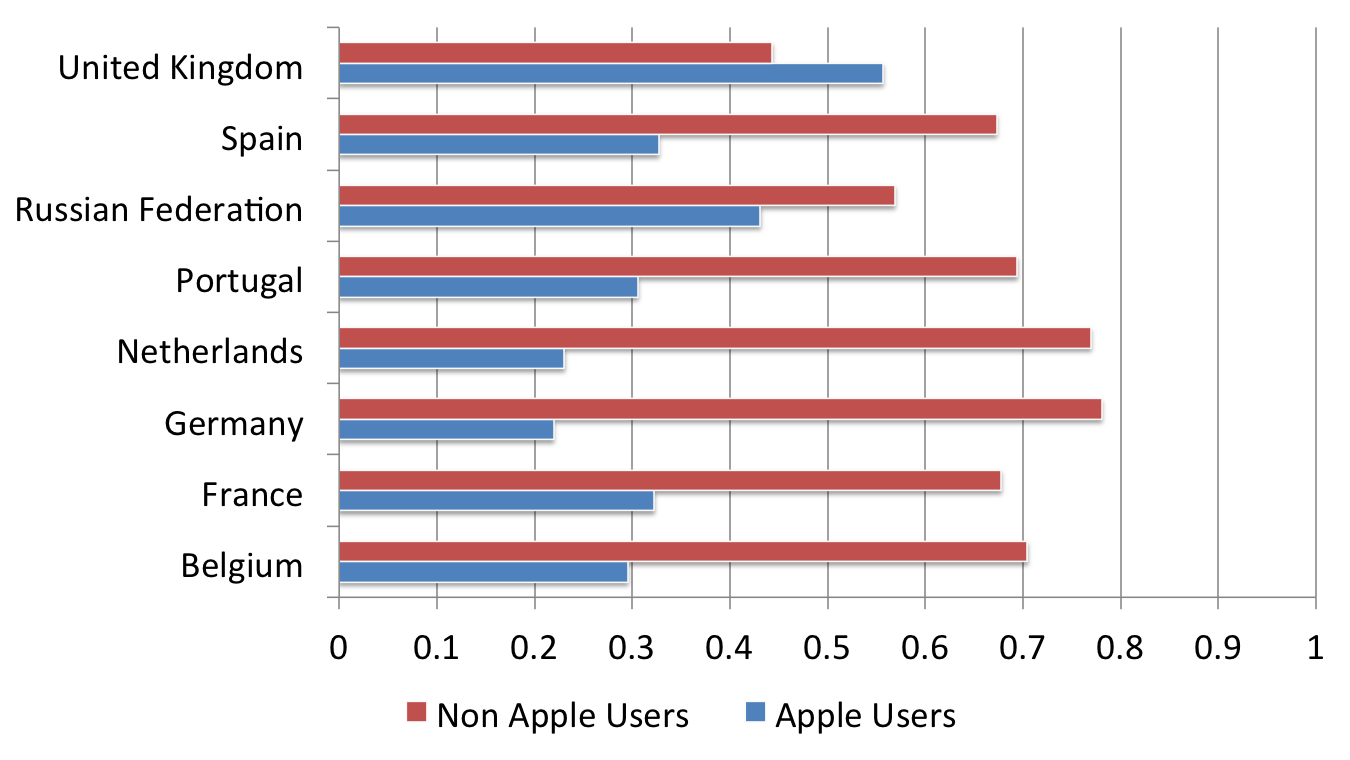}
	\caption{Phone type distribution across country}
	\label{phone_type}
\end{figure}

The brand, model, operation system of the phone can be obtained with IMEI-TAC code of the records. The price of the phone are scraped from Amazon API based on the above information. The phone type distribution across nations are shown in Figure \ref{phone_type}. 

Note that accurate revenues and individual spendings generated from events can be measured by banking transactions \cite{telefonica} if available.

%In addition, impact on hospitality can be measured by length of stay. We therefore estimate the stay length and night-stay of tourists as proxy for the influence on revenues for hotels. 

\subsection{Tourists interests profiles}
The understanding of tourists' interests offer businesses and tourism department a series of tactical and strategic recommendations on deicision-makings. As detailed footprints are tracked, the behavioral patterns for each interest category increase the knowledge of the industry's target market. This enables businesses to promote various service bundles to retain and attract customers. 

Personal interests are inferred in two steps. We first integrate the geo-location of cell towers and Point of Interests (POIs), tagging each cell towers with the potential activities individuals may perform nearby. Second, we aggregate the cell towers each individual connect to. With this, we obtain a distribution for activities individuals are interested in.

\section{Case Study}
\label{res}

In this section, we use the proposed indicators in section \ref{method} to showcase a comprehensive assessment of the high-impact events based on CDRs, extending beyond traditional indicators learned from surveys.

\subsection{Summer events analytics}
\begin{figure}[h]                                        	\includegraphics[width=\linewidth,natwidth=1500,natheight=836]{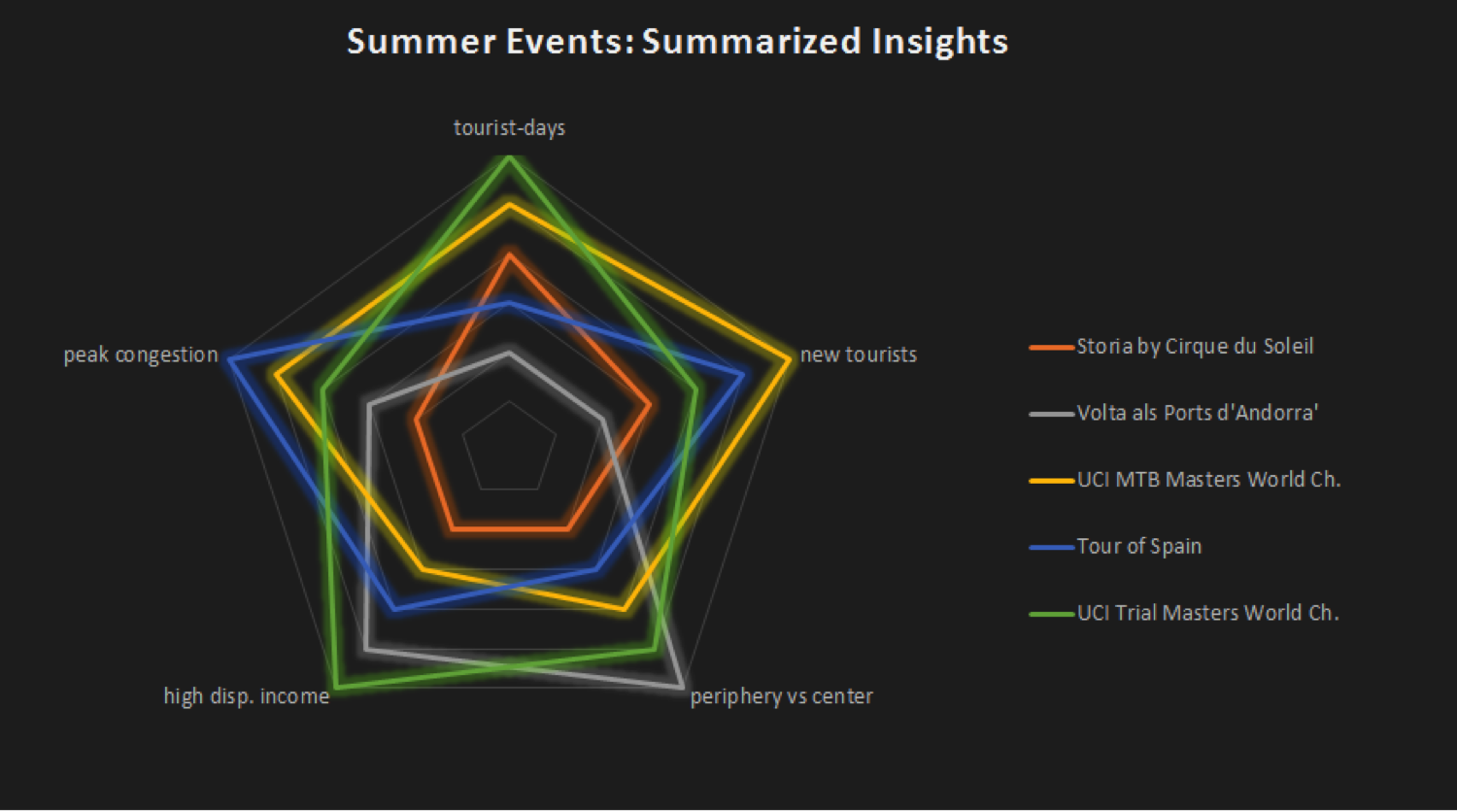}
	The relative position indicates the relative performance in the five events. Each spoke represents each indicator, with colors standing for events.
	\caption{Summer events analytics }
	\label{summer}
\end{figure}

Figure \ref{summer} shows the performances of five summer events on a radar chart, based on tourist days, new tourists, spatial distribution, income profiles and peak congestion. We target the following summer events: 

\begin{itemize} 
		\item Storia by Cirque du Soleil: 2015/07/15 - 2015/07/20.
		\item Volta als Ports d'Andorra: 2015/07/12. 
		\item MTB Masters World Championships: 2015/08/23 - 2015/08/27. 
		\item Tour of Spain: 2015/08/31 - 2015/09/03. 
		\item UCI Trial Masters World Championships: 2015/09/01 - 2015/09/06.
\end{itemize}

Insights from Figure \ref{summer} include: 
\begin{itemize}
\item UCI Trials Master World Championship was relatively more successful in most dimensions. Planners carried out successful marketing strategies and winned high popularity. Activities were not as attractive to high-income visitors as low-income ones, the former ones may generate more revenues. 
%\item Cirque De Soleil generated small spatial impact, smallest proportion of rich visitors and new tourists. 
\item MTB World Master Championships attracted a large volume of tourist and new tourists, indicating the popularity and success in marketing strategies. Tourists were centralized in central area and caused severe congestion, indicating more efforts in: 1) marketing suburban areas; 2) organize events or develop businesses in attracting high-income visitors; 3) manage travel demand to reduce congestion. 

\end{itemize}

\subsection{Winter events analytics}
Tourist flows during events measures the popularities of the events. However, more insights about the characteristics and segmentation of the population dynamics reveal better knowledge about marketing strategy prior to the events and travel experiences during the events. The targeted winter events include:

	\begin{itemize}
		\item Free-ride Junior World Championship: 2015/02/07 - 2015/02/08
		\item Speed Skiing World Championship: 2015/02/28 - 2015/03/03
		\item Alpine Skiing European Cup Finals: 2015/03/18 - 2015/3/22
		\item Total Fight Master of Freestyle: 2015/03/26 - 2015/04/04
	\end{itemize}

\begin{figure}[h]
	\includegraphics[width=\linewidth,natwidth=1069,natheight=615]{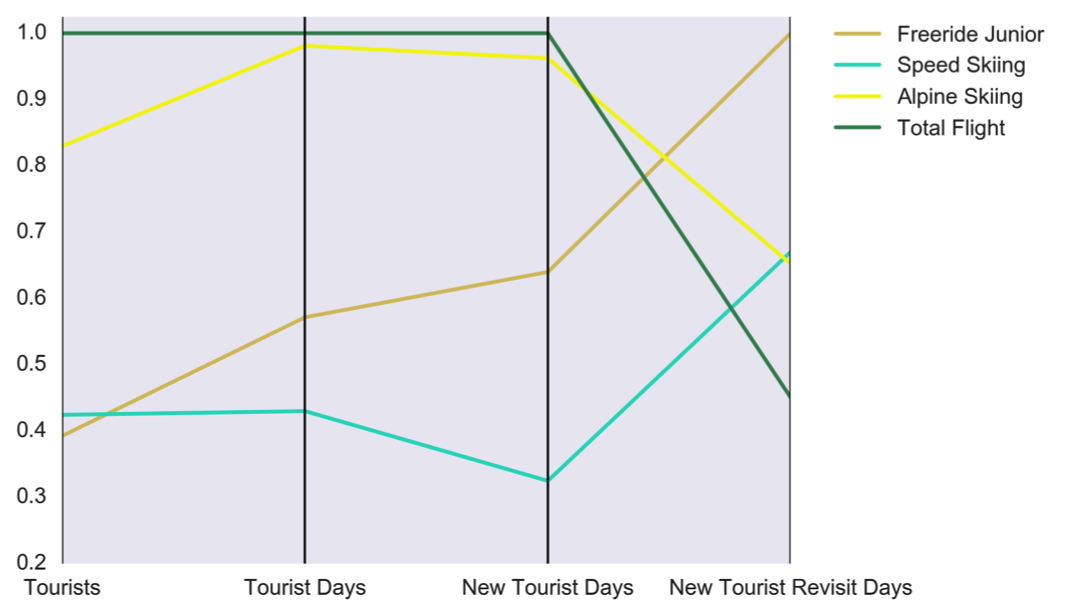}
	\caption{Winter events analytics }
	\label{winter}
\end{figure}

As shown in Figure \ref{winter}, Free-ride Junior World Championship, though less popular, attracted a large number of new tourists and repeated visits. This indicates its success in not only the marketing, but also the management and organization of the events. On the contrary, though Total Flight Master of Freestyl had high popularity, lowest revisit rate implies poor travel experiences, requiring more analytics on the specific aspects to be improved to retain tourists. 

\subsection{Tourists interests profiles}
\begin{figure}[h]
	\includegraphics[width=\linewidth,natwidth=2515,natheight=1656]{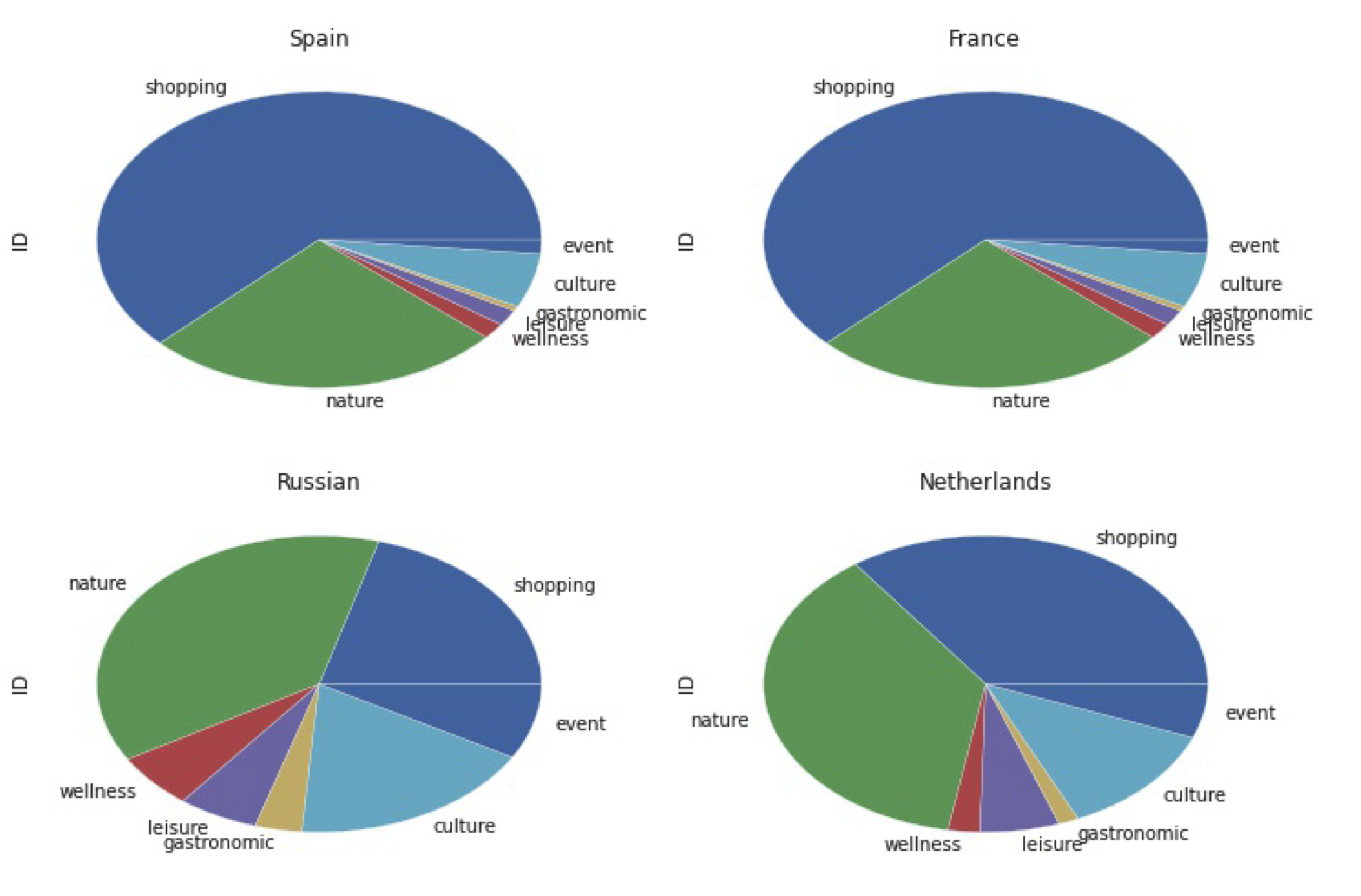}
	\caption{Nationality vs. interests}
	\label{interest}
\end{figure}

The interests of tourists across nations are informative in target marketing. As shown in Figure \ref{interest}, the Spanish and the French are more interested in shopping and nature activities, which take up $\frac{2}{3}$ and $\frac{1}{4}$ respectively. $\frac{2}{3}$ Russian are interested in nature activities, with another $\frac{1}{6}$ interested in cultural activities. $\frac{1}{3}$ Netherlands are interested in nature and shopping activities.

\section{Concluding remarks}
\label{future}

Growth in leisure travel has become increasingly significant economically, socially, and environmentally, accounting for a significant share of GDP and labor force. Policy and research on tourism traditionally rely on surveys and economic datasets, which are based on small samples and depict tourism dynamics at low spatial and temporal granularity. Anonymous call detail records (CDRs) are a novel source of data, with an enormous potential in areas of high societal value, such as epidemics, poverty, and urban development. 

This study demonstrates how to make use of the potential and added value of CDR for the formulation, analysis and evaluation of tourism strategies, at the national and local levels. In the context of the European country of Andorra, we use CDRs to evaluate marketing strategies, understand tourists' experiences, and evaluate revenues and externalities generated by touristic events. We do this by extracting many indicators in high spatial and temporal resolutions, such as tourist flows per country of origin, flows of new tourists, revisitation patterns, tourist externalities on transportation congestion, spatial distribution, economic impact, and profiling of tourist interests. Some of the indicators are traditionally used by tourism industry and others could not be quantified without large-scale and longitudinal data. We exemplify the use of these indicators for the planning and evaluation of high impact touristic events in Andorra, such as cultural festivals and sports competitions.

For future study, we plan to build an overreaching and automatic platform for tourisms and events management. The application of CDR data in tourism is promising in that it can be integrated with other geopositioned data for a comprehensive tourism monitoring and optimization platform, such as WiFi - more granular in movements, social media - momentary feelings and hidden interests, banking transactions - accurate revenue. 

\section{Acknowledgement}
This project is in collaboration with Changing Places Group at MIT Media lab, directed by Professor Kent Larson. We thank them, especially Arnaud Grignard, Naichun Chen, Nuria Macia and Yan Zhang, for the data support, visualizations and discussions.
\nocite{*}
\bibliographystyle{abbrv}
\bibliography{ref}

\end{document}